\documentclass[english,aps,prd,
groupedaddress,amssymb,
showpacs,nofootinbib]{revtex4}

\usepackage{times}
\usepackage{amsmath}
\usepackage{graphicx}

\newcommand{\beq}{\begin{eqnarray}}
\newcommand{\eeq}{\end{eqnarray}}

\makeatletter
\numberwithin{equation}{section}
\numberwithin{figure}{section}

\makeatother

\usepackage{babel}

\begin{document}
\setlength{\unitlength}{1mm}

\title{Longitudinal Rescaling of Quantum Electrodynamics}

\author{Axel \surname{Cort\'es Cubero}$^{1,2}$}

\email{acortes_cubero@gc.cuny.edu}

\author{Peter \surname{Orland}$^{1,2}$}

\email{orland@nbi.dk}

\affiliation{1. Baruch College, The 
City University of New York, 17 Lexington Avenue, 
New 
York, NY 10010, U.S.A. }

\affiliation{2. The Graduate School and University Center, The City University of New York, 365 Fifth Avenue,
New York, NY 10016, U.S.A.}



\begin{abstract}
We investigate quantum longitudinal rescaling of electrodynamics, transforming coordinates as $x^{0,3}\to\lambda x^{0,3}$ and $x^{1,2}\to x^{1,2}$, to one loop. We do this by an aspherical Wilsonian renormalization,  which was applied earlier to pure Yang-Mills theory. We find the anomalous powers of $\lambda$ in the renormalized couplings. Our result is only valid for 
$\lambda \lesssim 1$, because perturbation theory breaks down for $\lambda \ll 1$.
\end{abstract}


\maketitle

\section{Introduction} 
\setcounter{equation}{0}
\renewcommand{\theequation}{1.\arabic{equation}}

Field-theory actions greatly simplify under classical longitudinal rescaling, namely $x^{0,3}\to\lambda x^{0,3}$ and $x^{1,2}\to x^{1,2}$ 
\cite{Verlinde-squared}\cite{McLerranVenugopalan}. Such rescaling is interesting because the
center-of-mass energy transforms as $s\rightarrow \lambda^{-2}s$. Thus a large rescaling yields the high-energy limit. 

We consider the effect of longitudinal rescaling of quantum electrodynamics (QED) in this 
paper. The Abelian gauge field with Lorentz components $A_{\mu}$, $\mu=0,1,2,3$,
transforms as $A^{0,3}\to\lambda^{-1}A^{0,3}$ and $A^{1,2}\to A^{1,2}$. The action of the Maxwell gauge field is $S_{G}=-\frac{1}{4g^{2}}\int d^{4}xF_{\mu\nu}F^{\mu\nu}$, where $F_{\mu\nu}=\partial_{\mu}A_{\nu}-\partial_{\nu}A_{\mu}$, indices are raised with the usual Minkowski metric and $g$ is the bare electric charge.  Under longitudinal
rescaling, the gauge action becomes 
\beq
S_G\longrightarrow \frac{1}{4g^{2}}\int d^{4}x \left(F_{01}^{2}+F_{02}^{2}-F_{13}^{2}-F_{23}^{2}+\lambda^{-2}F_{03}^{2}-\lambda^{2}F_{12}^{2}\right).
\label{rescaledGauge}
\eeq
In the high-energy limit ($\lambda\to0$), the purely transverse terms may be neglected (this is easiest to see by an examination of the Hamiltonian). The massless Dirac action transforms as
 \beq
 S_{\rm Dirac}\longrightarrow i\int d^{4}x\,\bar{\psi}\left[\lambda^{-1}\gamma^0D_{0}+\lambda^{-1}\gamma^{3}D_{3}+\gamma^{1}D_{1}+\gamma^{2}D_{2}\right]\psi, \nonumber
 \eeq
where $D_{\mu} \psi=\partial_{\mu}\psi+iA_{\mu}\psi$ is the covariant derivative of $\psi$.
If we make an additional rescaling of  the spinor field $\psi\to\lambda^{\frac{1}{2}}\psi$ and $\bar{\psi}\to\lambda^{\frac{1}{2}}\bar{\psi}$, we obtain
 \beq
 S_{\rm Dirac}\longrightarrow i\int d^{4}x\,\bar{\psi}\left[\gamma^{0}D_0+\gamma^{3}D_3+\lambda\gamma^{1}D_1+\lambda\gamma^{2}D_2\right]\psi. \label{rescaledDirac}
 \eeq
In the quantum theory, anomalous powers of $\lambda$ will appear in the rescaled action
(\ref{rescaledGauge}) (\ref{rescaledDirac}). 

In the quantum theory, classical rescalings are no longer possible. The best-known is example is
the effect of a dilatation on a classically conformal invariant field theory. Quantum corrections violate this classical symmetry. 

An intuitive picture of quantum longitudinal rescaling is the following. Imagine cutting off the quantum field theory in the ultraviolet by a cubic lattice, with lattice spacing
$a$. Rescaling changes the lattice spacing of longitudinal coordinates to $\lambda a$, but does not change the lattice spacing of transverse coordinates, making the cut-off anisotropic. We therefore use a 
two-step process, where we first integrate out high-longitudinal momentum degrees of freedom, then restore isotropy with 
longitudinal rescaling. Instead of a lattice, we use a sharp-momentum cut-off and Wilson's renormalization procedure, to integrate out 
high-momentum modes \cite{WK}. This was done in reference \cite{PhysRevD80} for pure Yang-Mills theory. 

In the next section we review basic Wilsonian renormalization. In Section III we examine QED with a spherical momentum cut-off (this is just to show how the procedure works).  In Section IV we do the renormalization for aspherical cut-offs, which treats longitudinal momenta and transverse momenta differently. In Section V we find the quantum corrections to the QED action. We present our conclusions in the last section.

\section{Wilsonian Renormalization}
\setcounter{equation}{0}
\renewcommand{\theequation}{2.\arabic{equation}}

We Wick rotate to obtain the standard Euclidean
metric, so that the action is
\beq
S=\int d^{4}x (\frac{1}{4g^{2}}F_{\mu\nu}F^{\mu\nu}+i {\bar \psi}\,/\!\!\!\!D \,\psi). \nonumber
\eeq
where raising and lowering of indices is done with the Euclidean metric and where the slash on a vector quantity $J_{\mu}$ is $/\!\!\! J=\gamma^{\mu}J_{\mu}$, where 
$\gamma^{\mu}$ are the Euclidean Dirac matrices.

 We choose cut-offs $\Lambda$ and $\tilde \Lambda$ to be real 
positive numbers with units of
$cm^{-1}$ and $b$ and $\tilde b$ to be two dimensionless real numbers, such that
$b\ge 1$ and ${\tilde b}\ge 1$. These quantities satisfy $\Lambda>{\tilde \Lambda}$
and that $\Lambda^{2}/b \ge {\tilde \Lambda}^{2}/{\tilde b}$.  We  
introduce the ellipsoid in momentum space $\mathbb P$, which is the set of points $p$, such that
$bp_{L}^{2}+p_{\perp}^{2}<\Lambda^{2}$. We define the smaller ellipsoid ${\tilde {\mathbb P}}$ 
to
be the
set of points $p$, such that ${\tilde b}p_{L}^{2}+p_{\perp}^{2}<{\tilde \Lambda}^{2}$. Finally,
we define $\mathbb S$ to be the shell between the two ellipsoidal surfaces
 ${\mathbb S}={\mathbb P}-{\tilde{\mathbb P}}$. 

We split our fields into ``slow" and ``fast" pieces:
\beq
\psi(x)=\tilde{\psi}(x)+\varphi (x), \;
\bar{\psi}(x)= \tilde{\bar{\psi}}(x)+\bar{\varphi}(x),\;
A_{\mu}(x) =\tilde{A_{\mu}} (x)+a_{\mu} (x), \label{slow-fast}
\eeq
where the Fourier components of $\psi(x)$, ${\bar \psi}(x)$ and $A_{\mu}(x)$ vanish outside the
ellipsoid $\mathbb P$, the Fourier components of the slow fields  ${\tilde \psi}(x)$, ${\tilde{ \bar{ \psi}}}(x)$ and ${\tilde A}_{\mu}(x)$ vanish outside the
inner ellipsoid $\tilde {\mathbb P}$, and the Fourier components of the fast fields 
$\varphi(x)$, ${\bar \varphi}(x)$ and $a_{\mu}(x)$ vanish outside of the shell $\mathbb S$. Explicitly
\beq
\tilde{\psi}(x)=\int_{\tilde{\mathbb{P}}}\frac{d^{4}p}{(2\pi)^{4}}\psi(p)e^{-ip\cdot x}, \,\,\varphi(x)
\!&\!=\!&\!\int_{\mathbb{S}}\frac{d^{4}p}{(2\pi)^{4}}\psi(p)e^{-ip\cdot x},\;
\tilde{\bar{\psi}}(x)=\int_{\tilde{\mathbb{P}}}\frac{d^{4}p}{(2\pi)^{4}}\bar{\psi}(p)e^{ip\cdot x},\,\,\bar{\varphi}(x)=\int_{\mathbb{S}}\frac{d^{4}p}{(2\pi)^{4}}\bar\psi(p)e^{ip\cdot x}, \nonumber \\
\tilde{A}_{\mu}(x)\!&\!=\!&\!
\int_{\tilde{\mathbb{P}}}\frac{d^{4}p}{(2\pi)^{4}}A_{\mu}(p)e^{-ip\cdot x},\,\,\,a_{\mu}(x)=\int_{\mathbb{S}}\frac{d^{4}p}{2(\pi)^{4}}A_{\mu}(p)e^{-ip\cdot x}.\nonumber
\eeq
We denote the field strength of the slow fields by ${\tilde F}_{\mu\nu}=
\partial_{\mu}{\tilde A}_{\nu}-\partial_{\nu}{\tilde A}_{\mu}$. The functional integral with the ultraviolet cut-off $\Lambda$ and anisotropy parameter $b$ is
\beq
Z=\int_{\mathbb{P}}\;\mathcal{D}\psi\mathcal{D}\bar{\psi}\mathcal{D}A e^{-S} .\label{originalZ}
\eeq
There is no renormalization of a gauge-fixing parameter, because
we do not impose a gauge condition on the slow gauge field. We do impose a Feynman gauge condition
on the fast gauge field. As we show below, counterterms must be included in the action to maintain gauge invariance. We expect that renormalizability of the the field theory implies that these have the same form at each loop order; we have not proved this, however.

Before integrating over the fast fields, we 
must expand the action to second order in these fields. This expansion is
\beq
S={\tilde S}+S_{0}+S_{1}+S_{2}+S_{3},\nonumber
\eeq
where 
\beq
{\tilde S}\!&\!=\!&\!\int d^{4}x ({\tilde F}_{\mu\nu}{\tilde F}^{\mu\nu}+i {\tilde {\bar \psi}}\,/\!\!\!\!D \,{\tilde \psi}), \nonumber \\
S_{0}\!&\!=\!&\! \int_{\mathbb S} \frac{d^{4}q}{(2\pi)^{4}}\frac{1}{2}q^{2} a_{\mu}(-q) a^{\mu}(q)
+i\int_{\mathbb{S}}\frac{d^{4}q}{(2\pi)^{4}}\bar{\varphi}(-q)\; /\!\!\!q
\;\varphi(q), \nonumber \\
S_{1}\!&\!=\!&\!
-\int_{\mathbb{S}}\frac{d^{4}q}{(2\pi)^{4}}\int_{\tilde{\mathbb{P}}}\frac{d^{4}p}{(2\pi)^{4}}\left[\bar{\varphi}(q)/\!\!\!\!A(p)\varphi(-q-p)\right], \nonumber \\
S_{2} \!&\!=\!&\!  -\int_{\mathbb{S}}\frac{d^{4}q}{(2\pi)^{4}}\int_{\tilde{\mathbb{P}}}\frac{d^{4}p}{(2\pi)^{4}}\left[\bar{\psi}(p)/\!\!\!a(q)\varphi(-q-p)\right], \nonumber \\
S_{3}\!&\!=\!\!&\!\!-\int_{\mathbb{S}}\frac{d^{4}q}{(2\pi)^{4}}\int_{\tilde{\mathbb{P}}}\frac{d^{4}p}{(2\pi)^{4}}\left[\bar{\varphi}(-q-p)/\!\!\!a(q)\psi(p)\right]. \label{expansion}
\eeq
Notice that $S_2^*=S_3$. Henceforth, we drop the tildes on the slow fields, denoting these by 
$\psi,\, \bar{\psi}$ and $A_{\mu}$, but we keep the tilde on the slow action $\tilde S$.

The functional integral (\ref{originalZ}) may be written as
\beq
Z=\int_{\tilde{\mathbb{P}}}{\mathcal D}\;\psi{\mathcal D}\bar{\psi}{\mathcal D}A\;e^{-{\tilde S}}
\;\int_{\mathbb{S}}\;{\mathcal D}\varphi{\mathcal D}\bar{\varphi}{\mathcal D}a\;e^{-S_0}
e^{-S_{I}} \label{functionalintegral}
\eeq
where the integral of the interaction Lagranigian is $S_{I}=S_{1}+S_{2}+S_{3}$. To evaluate \ref{functionalintegral}, 
we use the fast-field propagators
\beq
\langle a_{\mu}(p)a_{\nu}(q)\rangle=\frac{g^{2}}{q^{2}}g_{\mu\nu}\delta^{(4)}(p+q)(2\pi)^{4}
\,,\;\;
\langle\varphi(p)\bar{\varphi}(q)\rangle=\frac{-i/\!\!\!q}{q^{2}}\delta^{(4)}(p+q)(2\pi)^{4}\,,
\label{propagators}
\eeq
where the brackets $\langle\,\rangle$ mean
\beq
\langle Q\rangle =\left(
\int_{\mathbb{S}}\;{\mathcal D}\varphi{\mathcal D}\bar{\varphi}{\mathcal D}a\;e^{-S_0}
\right)^{-1}\int_{\mathbb{S}}\;{\mathcal D}\varphi{\mathcal D}\bar{\varphi}{\mathcal D}a\;Q\;e^{-S_0}.
\label{bracket-def}
\eeq
for any quantity $Q$.
We will ignore an overall free-energy renormalization from the first factor in 
(\ref{bracket-def}). We use the connected-graph expansion
\beq
\langle e^{-S_{I}}\rangle=\exp [-\langle S_{I}\rangle+\frac{1}{2}(\langle S_{I}^{2}\rangle-\langle 
S_{I}\rangle^{2})-\frac{1}{3!}(\langle S_{I}^{3}\rangle-3\langle S_{I}^{2}\rangle\langle S_{I}\rangle+\langle S_{I}\rangle ^3)+\cdots], \label{cge}
\eeq
for the interaction $S_{I}$. The terms in the exponent of (\ref{cge}) are straightforward to evaluate using (\ref{propagators}). We find
\beq
\langle S_1\rangle=0\nonumber
\eeq
and
\beq
\frac{1}{2}\langle S_{1}^{2}\rangle=\int_{\tilde{\mathbb{P}}} \frac{d^{4}p}{(2\pi)^{4}}\;\Pi^{\mu\nu}(p)
A_{\mu}(p)A_{\nu}(-p), \nonumber
\eeq
where the polarization tensor $\Pi^{\mu\nu}(p)$ is defined as
\beq
\Pi^{\mu\nu}(p)=\frac{1}{2}\int_{\mathbb{S}}\frac{d^{4}q}{(2\pi)^{4}}{\rm Tr}\left[\frac{/\!\!\!q}{q^{2}}\gamma^\mu \frac{/\!\!\!q+/\!\!\!p}{(q+p)^{2}}\gamma^\nu\right].\label{polarizationtensor}
\eeq
Similarly
\beq
\langle S_{2}\rangle=\langle S_{3}\rangle=0\nonumber
\eeq
and
\beq
\langle S_{2}S_{3}\rangle=\langle S_{3}S_{2}\rangle=\int_{\mathbb{S}}\frac{d^{4}q}{(2\pi)^{4}}
\int_{\tilde{\mathbb{P}}}\frac{d^{4}p}{(2\pi)^{4}} \left[\frac{-(/\!\!\!p+/\!\!\!q)}{(p+q)^{2}}\gamma^\mu\frac{g^{2}}{q^{2}}\psi(p)\bar{\psi}(-p)\gamma_{\mu}\right]. \nonumber
\eeq
Thus
\beq
\frac{1}{2}(\langle S_{2}S_{3}\rangle+\langle S_{3}S_{2}\rangle)=\int_{\tilde{\mathbb{P}}}\frac{d^4p}{(2\pi)^4}\Sigma(p)\bar{\psi}(p)
\psi(-p),\nonumber
\eeq
where the self-energy correction $\Sigma(p)$ is
\beq
\Sigma(p)=g^{2}\int_{\mathbb{S}}\frac{d^{4}q}{(2\pi)^{4}}\left[\gamma^\mu\frac{i(/\!\!\!p+/\!\!\!q)}{(p+q)^{2}}\gamma_{\mu}\frac{1}{q^{2}}\right]=-2g^{2}\int_{\mathbb{S}}\frac{d^{4}q}{(2\pi)^{4}}\left[\frac{i(/\!\!\!p+/\!\!\!q)}{q^{2}(q+p)^{2}}\right].\label{selfenergy} 
\eeq

From the cubic term in (\ref{cge}), we find 
\beq
-\frac{1}{3!}(\langle S_{I}^{3}\rangle\!&\!-\!&\! 3\langle S_{I}^{2}\rangle\langle S_{I}\rangle+\langle S_{I}\rangle ^3)
=-\frac{1}{3!}\langle S_{I}^{3} \rangle 
=-\langle S_{1} S_{2}S_{3}\rangle \nonumber \\
&=&\int_{\tilde{\mathbb{P}}}\frac{d^4p}{(2\pi)^4}\int_{\tilde{\mathbb{P}}}\frac{d^4q}{(2\pi)^4}\bar{\psi}(p)\Gamma^{\mu}(p,q)A_{\mu}(q)\psi(-p-q),\nonumber
\eeq
where the vertex correction $\Gamma^\mu(p,q)$ is
\beq
\Gamma^\mu(p,q)=2g^{2}\int_{\mathbb{S}}\frac{d^{4}k}{(2\pi)^{4}}\frac{/\!\!\!k\gamma^\mu(/\!\!\!k+/\!\!\!q)}{(k-p)^{2}(k+q)^{2}k^{2}}. \label{vertex}
\eeq

\section{Spherical Momentum Cut-offs}
\setcounter{equation}{0}
\renewcommand{\theequation}{3.\arabic{equation}}

The cut-offs of our theory become isotropic if $b,\tilde{b}=1$. Then the region $\mathbb{P}$ is a sphere in momentum space, whose elements $p_\mu$ satisfy $p^{2}<\Lambda^2$. The region $\tilde{\mathbb{P}}$ is also a sphere, whose elements $q_\mu$ satisfy $q^2<\tilde{\Lambda}^2$. The region $\mathbb{S}$ is a spherical shell, $\mathbb{S}=\mathbb{P}-\tilde{\mathbb{P}}$.

The polarization tensor (\ref{polarizationtensor}) may be written
\beq
\Pi^{\mu\nu}(p)=tr\left[\frac{1}{2}\gamma^\mu\gamma^\alpha\gamma^\nu\gamma^\beta\int_{\mathbb{S}}\frac{d^{4}q}{(2\pi)^{4}}\frac{q_\alpha(q_\beta+p_\beta)}{q^{2}(q+p)^{2}}\right].\nonumber
\eeq
We expand the integrand in powers of $p$ to second order, to find
\beq
&\Pi^{\mu\nu}(p)=tr\left\{\frac{1}{2}\gamma^\mu\gamma^\alpha\gamma^\nu\gamma^\beta\frac{2\pi^{2}}{(2\pi)^{4}}\left[\int dq\frac{q^3\delta_{\alpha\beta}}{4q^{2}}-\int dq\frac{q^3p^{2}\delta_{\alpha\beta}}{4q^{4}}\right. \right.\nonumber\\
&\left.\left. -\int dq\frac{q^3p_\beta p^\gamma \delta_{\alpha\gamma}}{2q^{4}}+\int dq\frac{q^3}{6q^{4}}p^\gamma p^\delta(\delta_{\alpha\beta}\delta_{\gamma\delta}+\delta_{\alpha\delta}\delta_{\gamma\beta}+\delta_{\alpha\gamma}\delta_{\beta\delta})\right] \right\}.\label{ofour}
\eeq
To obtain (\ref{ofour}), we have used 
\beq
\int_{\mathbb{S}} d^{4}q \,q_\alpha q_\beta=2\pi^{2}\int_{\tilde{\Lambda}}^\Lambda dq\, q^3 \frac{\delta_{\alpha\beta} q^{2}}{4}\nonumber
\eeq
and
\beq
\int_{\mathbb{S}}d^{4}q\, q_\alpha q_\beta q_\gamma q_\delta=\frac{1}{24}\int_{\mathbb{S}} 
d^{4}q \,q^{4}(\delta_{\alpha\beta}\delta_{\gamma\delta}+\delta_{\alpha\delta}\delta_{\gamma\beta}+\delta_{\alpha\gamma}\delta_{\beta\delta}),\nonumber
\eeq
which follow from $\mathcal{O}(4)$ symmetry.
Thus the polarization tensor is
\beq
\Pi^{\mu\nu}(p)=tr\left\{\frac{1}{2}\gamma^\mu\gamma^\alpha\gamma^\nu\gamma^\beta\frac{1}{8\pi^{2}}\left[\frac{\delta_{\alpha\beta}}{8}(\Lambda^{2}-\tilde{\Lambda}^{2})-\frac{1}{12}(\delta_{\alpha\beta}p^{2}+2p_\alpha p_\beta)\ln \left(\frac{\Lambda}{\tilde{\Lambda}}\right)\right]\right\}.\label{logs}
\eeq
We must remove non-gauge-invariant terms, namely those quadratic in the cut-offs, with counterterms. The remaining logarithmically-divergent part of (\ref{logs}) is
\beq
&\hat\Pi^{\mu\nu}(p)=\Pi_{\mu\nu}(p)-\Pi_{\mu\nu}(0)= tr\left[\frac{1}{2}\gamma^\mu\gamma^\alpha\gamma^\nu\gamma^\beta \frac{1}{8\pi^{2}}\left(\frac{-1}{12}\right)(\delta_{\alpha\beta}p^{2}+2p_\alpha p_\beta)\ln \left(\frac{\Lambda}{\tilde{\Lambda}}\right)\right]\nonumber\\
&=\frac{e^{2}}{12\pi^{2}}(g^{\mu\nu}p^{2}-p^\mu p^\nu)\ln\left(\frac{\Lambda}{\tilde{\Lambda}}\right).\nonumber
\eeq
This gauge-invariant contribution satisfies $p^{\mu}\hat\Pi_{\mu\nu}(p)=0$. The contribution to the action associated with this term is
\beq
\frac{1}{2}\langle S_1^2\rangle=\int_{\tilde{\mathbb{P}}} \frac{d^4p}{(2\pi)^4}\,\frac{1}{12\pi^2}\ln\left(\frac{\Lambda}{\tilde{\Lambda}}\right)(g^{\mu\nu}p^2-p^\mu p^\nu)A_{\mu}(-p)A_{\nu}(p).
\label{S1squared}
\eeq
Equation (\ref{S1squared}) gives the effective coupling $\tilde{g}$ for the theory with cutoff 
$\tilde{\Lambda}$:
\beq
\frac{1}{4\tilde{g}^{2}}=\frac{1}{4g^2} +\frac{1}{12\pi^2}\ln\left(\frac{\Lambda}{\tilde{\Lambda}}\right),\nonumber
\eeq
which yields the standard result for the QED beta function:
\beq
\beta(\tilde{g})=\frac{\partial {\tilde{g}}}{\partial\ln ({\Lambda})}=\frac{g^3}{12\pi^2} . \nonumber
\eeq

The self-energy correction (\ref{selfenergy}) is 
\beq
\Sigma(p)=-2g^{2}\int_{\mathbb{S}}\frac{d^{4}q}{(2\pi)^{4}}\frac{(/\!\!\!q+/\!\!\!p)}{q^{2}(q+p)^{2}}.\nonumber
\eeq
We expand the integrand of $\Sigma(p)$ in powers of $p$, which gives
\beq
\Sigma(p)=-2g^{2}\frac{\gamma^\alpha}{8\pi^{2}}\int dq\left[\frac{-p_\alpha}{2q}+\frac{p_\alpha}{q}\right]
=-\frac{g^{2}/\!\!\!p}{8\pi^{2}}\ln \left(\frac{\Lambda}{\tilde{\Lambda}}\right). 
\eeq

The vertex correction (\ref{vertex}) is
\beq
\Gamma^\mu(p,q)=2g^{2}\gamma^\alpha\gamma^\mu\gamma^\beta\int_{\mathbb{S}}\frac{d^{4}k}{(2\pi)^{4}}\frac{k_\alpha(k_\beta+q_\beta)}{(k-p)^{2}(k+q)^{2}k^{2}}.\nonumber 
\eeq
Expanding the integrand in powers of $p$,
\beq
\Gamma^\mu(p,q)=2g^{2}\gamma^\alpha\gamma^\mu\gamma^\beta\int_{\mathbb{S}}\frac{d^{4}k}{(2\pi)^{4}}\left[\frac{\delta_{\alpha\beta}}{4k^{4}}-\frac{\delta_{\alpha\beta}p^{2}}{4k^6}-\frac{\delta_{\alpha\beta}q^{2}}{4k^6}+\frac{\delta_{\alpha\gamma}}{2k^6}(q_\beta p^\gamma-q_\beta q^\gamma)\right. \nonumber \\
\left.-\frac{\delta_{\alpha\beta}\delta_{\gamma\delta}+\delta_{\alpha\gamma}\delta_{\beta\delta}+\delta_{\alpha\delta}\delta_{\beta\gamma}}{6k^6}(q^\gamma p^\delta-q^\gamma q^\delta-p^\gamma p^\delta)\right].\nonumber 
\eeq
We retain only the divergent part of $\Gamma^{\mu}(p,q)$, namely the first term:
\beq
\Gamma^\mu(p,q)=2g^{2}\frac{\gamma^\alpha\gamma^\mu\gamma^\beta}{8\pi^{2}}\frac{\delta_{\alpha\beta}}{4}\ln\left(\frac{\Lambda}{\tilde{\Lambda}}\right)=\frac{-g^{2}\gamma^\mu}{8\pi^{2}}\ln\left(\frac{\Lambda}{\tilde{\Lambda}}\right). 
\eeq

\section{Ellipsoidal Cut-offs}
\setcounter{equation}{0}
\renewcommand{\theequation}{4.\arabic{equation}}

Next we consider the more general ellipsoidal case. The integration over $\mathbb{S}$ is done by changing variables from $q_{\mu}$ to two variables $u$ and $w$, with units of momentum squared, and two angles $\theta$ and $\phi$. These variables are defined by
\beq
q_1=\sqrt{u}\cos\theta,\,\,q_2=\sqrt{u}\sin\theta,\,\,q_3=\sqrt{w-u}\cos\phi,\,\,q_0=\sqrt{w-u}\sin\phi.
\nonumber
\eeq
The integration over these variables is
\beq
\int_{\mathbb{S}}d^{4}q=\frac{1}{4}\int_{0}^{2\pi}d\theta\int_{0}^{2\pi}d\phi\left[\int_0^{\tilde{\Lambda}^{2}}du\int_{\tilde{b}^{-1}\tilde{\Lambda}^{2}+(1-\tilde{b}^{-1})u}^{b^{-1}\Lambda^{2}+(1-b^{-1})u}dw+\int_{\tilde{\Lambda}^{2}}^{\Lambda^{2}}du\int_{u}^{b^{-1}\Lambda^{2}+(1-b^{-1})u}dw\right].\nonumber 
\eeq
We have a $\mathcal{O}(2)\times\mathcal{O}(2)$ symmetry, generated by the translations $\theta\to \theta+d\theta$ and $\phi\to\phi+d\phi$, rather than $\mathcal{O}(4)$ symmetry.

Our three corrections are expressed in terms of the integrals
\beq
&A_{\alpha\beta}=\int_{\mathbb{S}}\frac{d^{4}q}{(2\pi)^{4}}\frac{q_\alpha q_\beta}{q^{4}}, 
\nonumber \\
&B_{\alpha\beta}=\int_{\mathbb{S}}\frac{d^{4}q}{(2\pi)^{4}}\frac{q_\alpha q_\beta}{q^6},
\nonumber \\
&C_{\alpha\beta\gamma\delta}=\int_{\mathbb{S}}\frac{d^{4}q}{(2\pi)^{4}}\frac{q_\alpha q_\beta q_\gamma q_\delta}{q^8} \nonumber
\eeq
and
\beq
D=\int_{\mathbb{S}}\frac{d^{4}q}{(2\pi)^{4}}\frac{1}{q^{4}}.
\eeq
By inspection we write
\beq
&\Pi^{\mu\nu}(p)=tr\left[\frac{1}{2}\gamma^\mu\gamma^\alpha\gamma^\nu\gamma^\beta\left[A_{\alpha\beta}+4C_{\alpha\beta\gamma\delta}p^\gamma p^\delta-p^{2}B_{\alpha\beta}-2B_{\alpha\gamma}p_\beta p^\gamma\right]\right],
\nonumber \\
&\Sigma(p)=-2g^{2}\gamma^\alpha\left[-2B_{\alpha\gamma}p^\gamma+p_\alpha D\right]
\nonumber
\eeq
and
\beq
\Gamma^\mu=2g^{2}\gamma^\alpha\gamma^\mu\gamma^\beta B_{\alpha\beta}.
\nonumber 
\eeq

We use $C$ and $D$ to denote Lorentz indices taking only the values $1$ and $2$. We use
$\Omega$ and $\Xi$ to denote Lorentz indices taking only the values $3$ and $0$. The integration is straightforward, though tedious. We present only the results: 
\beq
&A_{CD}=\frac{\delta_{CD}}{32\pi^{2}}\Lambda^{2}\left[1+\frac{b}{(b-1)^{2}}(1-b+\ln b)\right]-\frac{\delta_{CD}}{32\pi^{2}}\tilde{\Lambda}^{2}\left[1+\frac{\tilde{b}}{(\tilde{b}-1)^{2}}(1-\tilde{b}+\ln \tilde{b})\right],
\nonumber \\
&A_{\Omega\Xi}=\frac{\delta_{\Omega\Xi}}{32\pi^{2}}\left[\Lambda^{2}\left(\frac{1}{b-1}-\frac{\ln b}{(b-1)^{2}}\right)-\tilde{\Lambda}^{2}\left(\frac{1}{b-1}-\frac{\ln \tilde{b}}{(\tilde{b}-1)^{2}}\right)\right],
\nonumber \\
&A_{C\Omega}=0,
\nonumber \\
&B_{CD}=\frac{\delta_{CD}}{32\pi^{2}}\ln\left(\frac{\Lambda}{\tilde{\Lambda}}\right)-\frac{\delta_{CD}}{64\pi^{2}}\left[\frac{b^{2}\ln b}{(b-1)^{2}}-\frac{b}{b-1}\right]+\frac{\delta_{CD}}{64\pi^{2}}\left[\frac{\tilde{b}^{2}\ln \tilde{b}}{(\tilde{b}-1)^{2}}-\frac{\tilde{b}}{\tilde{b}-1}\right],
\nonumber \\
&B_{\Omega\Xi}=\frac{\delta_{\Omega\Xi}}{32\pi^{2}}\ln\left(\frac{\Lambda}{\tilde{\Lambda}}\right)-\frac{\delta_{\Omega\Xi}}{64\pi^{2}}\left[\frac{b(b-2)\ln b}{(b-1)^{2}}+\frac{b}{b-1}\right]+\frac{\delta_{\Omega\Xi}}{64\pi^{2}}\left[\frac{\tilde{b}(\tilde{b}-2)\ln \tilde{b}}{(\tilde{b}-1)^{2}}+\frac{\tilde{b}}{\tilde{b}-1}\right],
\nonumber \\
&B_{C\Omega}=0,
\nonumber \\
&C_{CCCC}=\frac{1}{64\pi^{2}}\ln \left(\frac{\Lambda}{\tilde{\Lambda}}\right)-\frac{1}{128\pi^{2}}\frac{b^3}{(b-1)^3}\left(\ln b-\frac{2(b-1)}{b}+\frac{(b-1)(b+1)}{2b^{2}}\right)
\nonumber \\
&+\frac{1}{128\pi^{2}}\frac{\tilde{b}^3}{(\tilde{b}-1)^3}\left(\ln \tilde{b}-\frac{2(\tilde{b}-1)}{\tilde{b}}+\frac{(\tilde{b}-1)(\tilde{b}+1)}{2\tilde{b}}\right),\nonumber \\
&C_{1122}=\frac{C_{CCCC}}{3},
\nonumber\\
& C_{\Omega\Omega\Omega\Omega}=\frac{1}{64\pi^{2}}\ln \left(\frac{\Lambda}{\tilde{\Lambda}}\right)-\frac{1}{128\pi^{2}}\left[\frac{b^3}{(b-1)^3}\left(\ln b-\frac{2(b-1)}{b}+\frac{(b-1)(b+1)}{2b^{2}}\right)+\frac{3b\ln b}{b-1}-\frac{3b^{2}\ln b}{(b-1)^{2}}+\frac{3b}{b-1}\right]
\nonumber\\
&+\frac{1}{128\pi^{2}}\left[\frac{\tilde{b}^3}{(\tilde{b}-1)^3}\left(\ln \tilde{b}-\frac{2(\tilde{b}-1)}{\tilde{b}}+\frac{(\tilde{b}-1)(\tilde{b}+1)}{2\tilde{b}^{2}}\right)+\frac{3\tilde{b}\ln \tilde{b}}{\tilde{b}-1}-\frac{3\tilde{b}^{2}\ln \tilde{b}}{(\tilde{b}-1)^{2}}+\frac{3\tilde{b}}{\tilde{b}-1}\right],
\nonumber\\
&C_{0033}=\frac{C_{\Omega\Omega\Omega\Omega}}{3},
\nonumber\\
&C_{CC\Omega\Omega}=\frac{1}{192\pi^{2}}\ln \left(\frac{\Lambda}{\tilde{\Lambda}}\right)-\frac{1}{384\pi^{2}}\left[\frac{-2b^{2}\ln b}{(b-1)^3}+\frac{b^{2}\ln b+2b}{(b-1)^{2}}\right]+\frac{1}{384\pi^{2}}\left[\frac{-2\tilde{b}^{2}\ln \tilde{b}}{(\tilde{b}-1)^3}+\frac{\tilde{b}^{2}\ln \tilde{b}+2\tilde{b}}{(\tilde{b}-1)^{2}}\right],
\nonumber\\
&D=\frac{1}{8}\ln \left(\frac{\Lambda}{\tilde{\Lambda}}\right)-\frac{1}{16\pi^{2}}\left[\frac{b\ln b}{b-1}-\frac{\tilde{b}\ln \tilde{b}}{\tilde{b}-1}\right].\label{results}\eeq

Setting $b=\tilde{b}$ in (\ref{results}), we recover the results from the spherical integration done in Section III.
We simplify by setting $b=1$ and $\tilde{b}\approx 1$, using the expansion $\tilde{b}=1+\ln \tilde{b}+\frac{\ln^{2}\tilde{b}}{2}+\frac{\ln^3\tilde{b}}{3!}+\cdots$ and $\ln \tilde{b}=\ln \tilde{b}-\frac{\ln^{2}\tilde{b}}{2}+\frac{\ln^3 \tilde{b}}{3}-\frac{\ln^{4} \tilde{b}}{4}+\cdots $, dropping terms of second order in $\ln \tilde{b}$.

The vertex correction is
\beq
\Gamma^\mu(p,q)=2g^{2}\gamma^\alpha\gamma^\mu\gamma^\beta B_{\alpha\beta}=2g^{2}\left(\frac{-\gamma^\mu}{16\pi^{2}}\ln\left(\frac{\Lambda}{\tilde{\Lambda}}\right)-\frac{\gamma^\mu}{16\pi^{2}}\ln\tilde{b}+\frac{g^{C\mu}\gamma_C}{32\pi^{2}}\frac{5}{6}\ln\tilde{b}+\frac{g^{\Omega\mu}\gamma_{\Omega}}{32\pi^{2}}\frac{7}{6}\ln\tilde{b}\right).
\eeq

The self-energy correction is
\beq
\Sigma(p)=-2g^{2}\gamma^\alpha[-2B_{\alpha\beta}p^\beta+p_\alpha D]=-2g^{2}\left[\frac{\gamma^\mu p_{\mu}}{16\pi^{2}}\ln \left(\frac{\Lambda}{\tilde{\Lambda}}\right)+\frac{1}{32\pi^{2}}\ln \tilde{b}\frac{\gamma^C p_C}{6}-\frac{1}{32\pi^{2}}\ln \tilde{b}\frac{\gamma^\Omega p_\Omega}{6}\right].
\eeq
In the next section, we show how these affect the effective action.


The most general gauge-field action which is quadratic in $A_{\mu}$, is $\mathcal{O}(2)\times\mathcal{O}(2)$ invariant and gauge invariant, to leading order is
\beq 
S_{\rm quadratic}=\int_{\mathbb{P}}\frac{d^{4}p}{(2\pi)^{4}}A(-p)^T\left[a_{1}M_1(p)+a_{2}M_2(p)+a_{3}M_3(p)\right]A(p),
\nonumber
\eeq
where
\beq
 M_1(p)=\left(\begin{array}{cccc}p_2^{2}&-p_1p_2&0&0\\\noalign{\medskip}-p_1p_2&p_1^{2}&0&0\\\noalign{\medskip}0&0&0&0\\0&0&0&0\end{array}\right),
M_2(p)=\left(\begin{array}{cccc}0&0&0&0\\0&0&0&0\\\noalign{\medskip}0&0&p_0^{2}&-p_3p_0\\\noalign{\medskip}0&0&-p_3p_0&p_3^{2}\end{array}\right),
\nonumber
\eeq
\beq 
M_3(p)=\left(\begin{array}{cccc}p_L^{2}&0&-p_1p_3&-p_1p_0\\\noalign{\medskip}0&p_L^{2}&-p_2p_3&-p_2p_0\\\noalign{\medskip}-p_1p_3&-p_2p_3&p_{\perp}^{2}&0\\\noalign{\medskip}-p_1p_0&-p_2p_0&0&p_{\perp}^{2}\end{array}\right),
\nonumber
\eeq
and $a_{1},\,a_{2}$ and $a_{3}$ are real numbers.
Any part of the polarization tensor that cannot be expressed in terms of these matrices (i.e. $\int_{\tilde{\mathbb{P}}}\frac{d^4p}{(2\pi)^4}A_\mu(-p)\Pi_{\mu\nu}(p)A_{\nu}(p)-S_{\rm quadratic}$) must be removed with counterterms. After some work we find
\beq
&\Pi^{\mu\nu}(p)=tr\left[\frac{1}{2}\gamma^\mu\gamma^\alpha\gamma^\nu\gamma^\beta\left[A_{\alpha\beta}+4C_{\alpha\beta\gamma\delta}p^\gamma p^\delta-p^{2}B_{\alpha\beta}-2B_{\alpha\gamma}p_\beta p^\gamma\right]\right]
\nonumber\\
&=\frac{1}{12\pi^{2}}\ln\left(\frac{\Lambda}{\tilde{\Lambda}}\right)(p^{2}\mathbf{1}-pp^T)^{\mu\nu}+\frac{5\ln\tilde{b}}{48\pi^{2}}(p^{2}\mathbf{1}-pp^T)^{\mu\nu}
\nonumber\\
&+\frac{\ln\tilde{b}}{128\pi^{2}}\left[\frac{8}{9}M_3+\frac{40}{9}M_2-\frac{104}{9}M_1+\frac{8}{3}\left(\begin{array}{cc}\left(\frac{17}{6}p_\perp^{2}+\frac{4}{3}p_L^{2}\right)\mathbf{1}_{2\times2}&0\\0&-\left(\frac{7}{6}p_L^{2}+\frac{14}{3}p_\perp^{2}\right)\mathbf{1}_{2\times2}\end{array}\right)\right]^{\mu\nu}.\label{polarization}
\eeq
This determines $a_{1},\,a_{2}$ and $a_{3}$, so that
\beq 
S_{\rm diff}=\int_{\tilde{\mathbb{P}}}\frac{d^{4}p}{(2\pi^{2})}A(-p)^TM_{\rm diff}A(p)=\int_{\tilde{\mathbb{P}}}\frac{d^{4}p}{(2\pi)^{2}}A(-p)^T \Pi A(p)-S_{\rm quadratic}
\eeq
is maximally non-gauge invariant. The matrix $M_{\rm diff}$ is the last diagonal matrix in (\ref{polarization}). The quantity $S_{\rm diff}$ is proportional to the local counterterms to include 
in the action. We find
\beq
a_{1}=\frac{1}{12\pi^2}\ln\left(\frac{\Lambda}{\tilde{\Lambda}}\right)+\left(\frac{5}{48\pi^2}-\frac{1}{128\pi^{2}}\frac{104}{9}\right)\ln\tilde{b},\,\,a_{2}=\frac{1}{12\pi^{2}}\ln\left(\frac{\Lambda}{\tilde{\Lambda}}\right)+\left(\frac{5}{48\pi^{2}}+\frac{1}{128\pi^{2}}\frac{40}{9}\right)\ln\tilde{b},\nonumber
\eeq
and
\beq
a_{3}=\frac{1}{12\pi^{2}}\ln\left(\frac{\Lambda}{\tilde{\Lambda}}\right)+\left(\frac{5}{48\pi^{2}}+\frac{1}{128\pi^{2}}\frac{8}{9}\right)\ln\tilde{b}\nonumber
\eeq
In the next section, we show how the action changes under renormalization. We then rescale to restore the isotropy.

\section{The Rescaled Effective Action}
\setcounter{equation}{0}
\renewcommand{\theequation}{5.\arabic{equation}}

We define the effective action $S'$, which contains the effects of integrating out the fast fields, by
\beq
Z=\int_{\tilde{\mathbb{P}}}\mathcal{D}\psi\mathcal{D}\bar{\psi}\mathcal{D}A\,e^{-S'}=\int_{\tilde{\mathbb{P}}}\mathcal{D}\psi\mathcal{D}\bar{\psi}\mathcal{D}A\,e^{-\tilde{S}}\;\int_{\mathbb{S}}\;{\mathcal D}\varphi{\mathcal D}\bar{\varphi}{\mathcal D}a\,e^{-S_0}
e^{-R},\nonumber
\eeq
where $S'=\int d^4x \left[\mathcal{L}_{\rm Fermion}+\mathcal{L}_{\rm vertex}+\mathcal{L}_{\rm gauge}\right]=\int d^4x\left[\mathcal{L}_{\rm Dirac}+\mathcal{L}_{\rm gauge}\right]$. To one loop
\beq
&\mathcal{L}_{\rm Fermion}=\bar{\psi}i (/\!\!\!\partial-\Sigma(\partial))\psi,\nonumber\\
&\mathcal{L}_{\rm vertex}=\bar{\psi}(\gamma^\mu-\Gamma^\mu)A_\mu\psi \nonumber
\eeq
and
\beq
\mathcal{L}_{\rm gauge}=\frac{1}{4g^2}F_{\mu\nu}F^{\mu\nu}+A_{\mu} (\sum_{i=1}^{3}a_{i}M_{i}^{\mu\nu}(\partial))A_\nu.\nonumber
\eeq 
Explicitly, $\mathcal{L}_{\rm vertex}$ is
\beq
&\mathcal{L}_{\rm vertex}=\bar{\psi}\left[\gamma^C\left(1+\frac{g^{2}}{8\pi^{2}}\ln\left(\frac{\Lambda}{\tilde{\Lambda}}\right)+\frac{g^{2}}{8\pi^{2}}\ln\tilde{b}-\frac{5g^{2}}{96\pi^{2}}\ln\tilde{b}\right)A_C\right.\nonumber\\
&\left.+\gamma^\Omega\left(1+\frac{g^{2}}{8\pi^{2}}\ln\left(\frac{\Lambda}{\tilde{\Lambda}}\right)+\frac{g^{2}}{8\pi^{2}}\ln\tilde{b}-\frac{7g^{2}}{96\pi^{2}}\ln\tilde{b}\right)A_\Omega\right]\psi\nonumber\\
&=R\bar{\psi}\left[\gamma^CA_C+\lambda^{\frac{g^{2}}{24\pi^{2}\tilde{R}}}\gamma^\Omega A_\Omega\right]\psi,\nonumber
\eeq
where 
\beq
R=\tilde{R}+\left(\frac{g^{2}}{8\pi^{2}}-\frac{5g^{2}}{96\pi^{2}}\right)\ln\tilde{b}\approx \tilde{R}\tilde{b}^{\frac{7g^{2}}{96\pi^{2}\tilde{R}}}=\tilde{R}\lambda^{-\frac{7g^{2}}{48\pi^{2}\tilde{R}}},\nonumber
\eeq
and
\beq
\tilde{R}=1+\frac{g^{2}}{8\pi^{2}}\ln\left(\frac{\Lambda}{\tilde{\Lambda}}\right),\nonumber
\eeq
for small $\ln\tilde{b}$, where we have identified $\tilde{b}=\lambda^{-2}$.

The term $\mathcal{L}_{\rm Fermion}$, which contains the self-energy correction:
\beq
&\mathcal{L}_{\rm Fermion}=\bar{\psi}i\left[\gamma^C\partial_C\left(1+\frac{g^{2}}{8\pi^{2}}\ln\left(\frac{\Lambda}{\tilde{\Lambda}}\right)+\frac{g^{2}}{8\pi^{2}}\ln\tilde{b}-\frac{5g^{2}}{96\pi^{2}}\ln\tilde{b}-\frac{g^{2}}{16\pi^{2}}\ln\tilde{b}\right)\right.\nonumber\\
&\left.+\gamma^\Omega\partial_\Omega\left(1+\frac{g^{2}}{8\pi^{2}}\ln\left(\frac{\Lambda}{\tilde{\Lambda}}\right)+\frac{g^{2}}{8\pi^{2}}\ln\tilde{b}-\frac{5g^{2}}{96\pi^{2}}\ln\tilde{b}-\frac{g^{2}}{12\pi^{2}}\ln\tilde{b}\right)\right]\psi\nonumber\\
&=R'\bar{\psi}i\left[\gamma^C\partial_C+\lambda^{\frac{g^{2}}{24\pi^{2}\tilde{R}}}\gamma^\Omega\partial_\Omega\right]\psi,\nonumber
\eeq
where 
\beq 
R'=R\tilde{b}^{-\frac{g^{2}}{16\pi^{2}\tilde{R}}}=R\lambda^{\frac{g^{2}}{8\pi^{2}\tilde{R}}}.\nonumber
\eeq
For consistency, we write $\mathcal{L}_{\rm vertex}$ in terms of $R'$,
\beq
\mathcal{L}_{\rm vertex}=R'\lambda^{-\frac{g^{2}}{8\pi^{2}\tilde{R}}}\bar{\psi}\left[\gamma^C A_C+\lambda^{\frac{g^{2}}{24\pi^{2}\tilde{R}}}\gamma^\Omega A_\Omega\right]\psi.\nonumber
\eeq

We must rescale the gauge field by
\beq
\lambda^{\frac{-g^{2}}{8\pi^{2}\tilde{R}}}A_{\mu}\to A_{\mu},\label{Arescaling}
\eeq
to express $\mathcal{L}_{\rm Dirac}=\mathcal{L}_{\rm Fermion}+\mathcal{L}_{\rm vertex}$ in terms of a covariant derivative.
This rescaling also affects $\mathcal{L}_{\rm gauge}$. We now have
\beq
\mathcal{L}_{\rm Dirac}=R'\bar{\psi}i\left[\gamma^C D_C+\lambda^{\frac{g^{2}}{24\pi^{2}\tilde{R}}}\gamma^\Omega D_\Omega\right]\psi.\nonumber
\eeq
Rescaling the spinor field by
\beq 
R'\lambda^{-1+\frac{g^{2}}{24\pi^{2}\tilde{R}}}\bar{\psi}\psi\to\bar{\psi}\psi,\nonumber
\eeq
gives us the form
\beq
\mathcal{L}_{\rm Dirac}=\bar{\psi}i\left[\lambda^{1-\frac{g^{2}}{24\pi^{2}\tilde{R}}}\gamma^CD_C+\gamma^\Omega D_\Omega\right]\psi.
\eeq

Including vacuum-polarization corrections, $\mathcal{L}_{\rm gauge}$ becomes
\beq
&\mathcal{L}_{\rm gauge}=\left(\frac{1}{4g^{2}}+\frac{1}{12\pi^{2}}\ln\left(\frac{\Lambda}{\tilde{\Lambda}}\right)+\frac{1}{9\pi^{2}}\ln\tilde{b}\right)(F_{01}^{2}+F_{02}^{2}+F_{13}^{2}+F_{23}^{2})\nonumber\\
&+\left(\frac{1}{4g^{2}}+\frac{1}{12\pi^{2}}\ln\left(\frac{\Lambda}{\tilde{\Lambda}}\right)+\frac{1}{9\pi^{2}}\ln\tilde{b}+\frac{1}{36\pi^{2}}\ln\tilde{b}\right)F_{03}^{2}\nonumber\\
&+\left(\frac{1}{4g^{2}}+\frac{1}{12\pi^{2}}\ln\left(\frac{\Lambda}{\tilde{\Lambda}}\right)+\frac{1}{9\pi^{2}}\ln\tilde{b}-\frac{7}{72\pi^{2}}\ln\tilde{b}\right)F_{12}^{2}.\nonumber
\eeq
We introduce the effective coupling $g_{\rm eff}$,
\beq
\frac{1}{g_{\rm eff}^{2}}=\frac{1}{\tilde{g}^{2}}+\frac{4}{9\pi^{2}}\ln\tilde{b}\approx \frac{1}{\tilde{g}^{2}}\tilde{b}^{\frac{4\tilde{g}^{2}}{9\pi^{2}}}=\frac{1}{\tilde{g}^{2}}\lambda^{\frac{-8}{9\pi^{2}}\tilde{g}^{2}},\nonumber
\eeq
where 
\beq
\frac{1}{\tilde{g}^{2}}=\frac{1}{g^{2}}+\frac{1}{3\pi^{2}}\ln\left(\frac{\Lambda}{\tilde{\Lambda}}\right).
\nonumber\\
\eeq
Then
\beq
\mathcal{L}_{\rm gauge}=\frac{1}{4g_{\rm eff}^{2}}\left(F_{01}^{2}+F_{02}^{2}+F_{13}^{2}+F_{23}^{2}+\lambda^{-\frac{2}{9\pi^{2}}\tilde{g}^{2}}F_{03}^{2}+\lambda^{\frac{7}{9\pi^{2}}\tilde{g}^{2}}F_{12}^{2}\right).\nonumber
\eeq
We finally rescale the gauge field with the factor from (\ref{Arescaling}), 
\beq
 F_{\mu\nu}^{2}\to\lambda^{\frac{g^{2}}{4\pi^{2}\tilde{R}}}F_{\mu\nu}^{2},\nonumber
\eeq
and define a new effective coupling $g'_{\rm eff}$ that absorbs this factor
\beq
\frac{1}{{g'}_{\rm eff}^{2}}=\frac{1}{\tilde{g}^{2}}\lambda^{-\frac{8}{9\pi^{2}}\tilde{g}^{2}+\frac{g^{2}}{4\pi^{2}\tilde{R}}},\label{effectivecoupling}
\eeq
\beq
\mathcal{L}_{\rm gauge}=\frac{1}{4{g'}_{\rm eff}^{2}}\left(F_{01}^{2}+F_{02}^{2}+F_{13}^{2}+F_{23}^{2}+\lambda^{-\frac{2}{9\pi^{2}}\tilde{g}^{2}}F_{03}^{2}+\lambda^{\frac{7}{9\pi^{2}}\tilde{g}^{2}}F_{12}^{2}\right).\eeq

Our final result, after longitudinal rescaling and Wick-rotating back to Minkowski space-time is
\beq
&\mathcal{L}=\mathcal{L}_{\rm Dirac}+\mathcal{L}_{\rm gauge}=\bar{\psi}i\left[\lambda^{1-\frac{g^{2}}{12\pi^{2}\tilde{R}}}\gamma^CD_C+\gamma^\Omega D_\Omega\right]\psi \nonumber \\
&+\frac{1}{4{g'}_{\rm eff}^{2}}\left(F_{01}^{2}+F_{02}^{2}-F_{13}^{2}-F_{23}^{2}+\lambda^{-2-\frac{2}{9\pi^{2}}\tilde{g}^{2}}F_{03}^{2}-\lambda^{2+\frac{7}{9\pi^{2}}\tilde{g}^{2}}F_{12}^{2}\right).\label{finalaction}
\eeq

\section{Conclusions}
\setcounter{equation}{0}
\renewcommand{\theequation}{6.\arabic{equation}}

In comparing our final action (\ref{finalaction}) to its classically-rescaled counterpart 
((\ref{rescaledGauge}) and (\ref{rescaledDirac})) we notice anomalous powers of $\lambda$, as well as corrections to the coupling constant (\ref{effectivecoupling}). As we take 
$\lambda\to 0$, the effective coupling ${g'}_{\rm eff}$ becomes weaker, as opposed to the growing coupling constant of QCD \cite{PhysRevD80}. The high-energy limit presents a problem for the effective transverse electric charge, which describes how transverse 
components of the gauge field are coupled to the Dirac field. The transverse electric charge 
$e_{\perp}$ is given by the coefficient of $F_{12}^2$ in the action, which we have called 
$\frac{1}{4e_{\perp}^2}$, and is 
\beq
e_{\perp}=\tilde{g}\lambda^{-1+\frac{\tilde{g}^{2}}{18\pi^{2}}-\frac{g^2}{8\pi^2\tilde{R}}}.\label{last}
\eeq
This coupling diverges even in the classical theory (note the factor $\lambda^{-1}$in 
(\ref{last})). The divergence is enhanced by the quantum correction. 
Our final action (\ref{finalaction}) is strictly valid only for $\lambda\lesssim 1$. Naively substituting 
$\lambda<< 1$ in our result suggest what the real high-energy action should be, though we 
cannot tell if this form is truly correct. We can extrapolate our results to higher energies, provided our couplings (particularly $e_{\perp}$) do not become large enough to invalidate 
perturbation theory. Naively, a Landau pole is encountered by solving our renormalization grouo equations at sufficiently high energy (the Landau pole itself is not meaningful, but is
just a signal that extrapolating to arbitrarily high energies is impossible).

It is interesting to see the role that fermions play in longitudinal rescaling. Pure Yang-Mills theory has been studied in Reference \cite{PhysRevD80}, where 
anomalous powers of the rescaling factor $\lambda$ were found. The quantum corrections due to fermions have the opposite effect to corrections due to the non-Abelian gauge fields (the anomalous powers of $\lambda$ that arise have opposite signs). We intend to study how QCD behaves with the inclusion of quarks. Particularly, we wish to see how anomalous dimensions depend on the number of flavors.

Eventually we wish to study longitudinal rescaling with a manifestly gauge invariant, but 
anisotropic, regularization (such as some version of dimensional regularization). It may then be possible to use the background-field method instead of 
Wilsonian renormalization. 


\begin{acknowledgments}

We thank Jamal Jalilian-Marian for discussions. This research was supported in
part by the National Science Foundation, under Grant No. PHY0855387
and by a grant from the PSC-CUNY.

\end{acknowledgments}

\end{document}